\documentclass[reprint,english,aps,pre,floatfix,showpacs]{revtex4-1}

\usepackage{amsmath}
\usepackage{graphicx}
\usepackage{amssymb}
\usepackage{babel}
\usepackage{color}
\usepackage{bm}
\renewcommand{\vec}[1]{\boldsymbol{#1} }

\begin{document}

\title{Role of defects in the onset of wall-induced granular convection}

\author{Andrea Fortini$^{\rm 1, 2}$}
\email{andrea.fortini@uni-bayreuth.de}

\author{Kai Huang$^{\rm 3}$}
\affiliation{$^{\rm 1}$Theoretische Physik II, Physikalisches Institut, Universit\"at Bayreuth, Universit\"atsstra{\ss}e 30, D-95447 Bayreuth,
Germany}
\affiliation{$^{\rm 2}$Department of Physics, University of Surrey, Guildford GU2 7XH, United Kingdom}

\affiliation{$^{\rm 3}$Experimentalphysik V, Physikalisches Institut, Universit\"at Bayreuth, Universit\"atsstra{\ss}e 30, D-95447 Bayreuth,
Germany}

\begin{abstract}
We investigate  the onset of  wall-induced convection in vertically vibrated granular matter by means of experiments and two-dimensional computer simulations. 
In both simulations and experiments we find that the wall-induced  convection occurs inside the \emph{bouncing bed} region of the parameter space in which the granular bed behaves like a bouncing ball. 
A good  agreement  between  experiments and simulations is found for the peak vibration acceleration at which convection starts.  
By comparing the results of simulations initialised with and without defects, we find that the onset of convection occurs at lower vibration strengths in the presence of defects. Furthermore, we find that  the convection of granular particles initialised in a perfect hexagonal lattice is related to the nucleation of defects and the process is described by an  Arrhenius law.
\end{abstract}

\maketitle

\section{Introduction}

Mixing and demixing of vibrated  granular matter~\cite{RevModPhys,luding2} are of importance in nature~\cite{Miyamoto:2007il} as well as in many industrial processes.  For example, they are used in the pharmaceutical, construction~\cite{Duran2000} or waste reprocessing~\cite{Mohabuth:2005eq} industries.
The term Brazil Nut Effect (BNE)~\cite{Rosato:1987dv}, which originally referred to the rise of a large  particle to the top of a container filled with smaller grains, is now used to indicate the more general demixing of differently sized particles under vertical oscillations.  \citet{schroter:2006cc} reviewed and identified seven possible mechanisms that  lead to the BNE and to the Reverse Brazil Nut Effect (RBNE)~\cite{Hong:2001kn}. 
Convective cells induced by the walls of the container were  found  to be a major  contributing mechanism to the occurrence of the BNE~\cite{Knight:1993bg,Cooke:1996gg,Poschel,Kudrolli:2004kr,Majid:2009jl}. 

The complexity of the BNE characterisation is in part due to an underlying dynamical behaviour which even for one component systems is very rich.
At driving accelerations smaller than the gravitational acceleration, the granular bed comoves with the bottom wall. Upon increasing the acceleration the granular bed behaves like a bouncing ball~\cite{Mehta:1990eh}, and above this bouncing bed region collective undulations (also known as arches)  appear~\cite{Douady:1989tc,Ugawa:2003im,Sano:2005ew,Eshuis:980854}. At still higher accelerations the granular \emph{Leidenfrost} effect occurs, in which a dense granular fluid hovers over a granular gas. Recently, \citet{Eshuis:2010ij} systematically drew  phase diagrams for all these phenomena, and found at very high accelerations  a convective regime in which the sample is completely fluidized~\cite{Eshuis:980854,Eshuis:2010ij}. 

However, this convection regime is distinct from  the wall-induced convection which occurs at low accelerations and is one of the driving mechanisms for the BNE. 
The wall-induced convection is caused by the shear forces between particles and walls.  During the upward acceleration the mixture gets compacted and shear forces induced by the side walls propagate efficiently through the whole sample. During the downward motion the mixture is more expanded and consequently those particles adjacent to the walls experience stronger downward shear forces than those in the centre of the container. The combination of the two type of motions gives rise to  convection~\cite{schroter:2006cc}. 
This cycle of expansion and compression of the granular bed was studied by~\citet{Sun:2006fy} and found to be strongly dependent on wall friction. Even though their study was done in relation to BNE,  no connection to the  convective motion is made.
The onset of convection has been studied extensively in both experiments~\cite{Clement92,knight} and with numerical simulations~\cite{taguchi,luding1,bourz,risso}.

In this article, we study with both computer simulations and experiments the dynamical phase diagram of two-dimensional vertically oscillated granular matter and analyse the mechanism behind the onset of the wall-induced convection to clarify the role of topological defects.

\section{Methods}

For the theoretical investigation we  carry out  Molecular Dynamic(MD)~\cite{Frenkel} simulations of a two-dimensional system in a box of size $L_x \times L_z$ delimited by flat hard walls and gravity $\vec g= -g \vec e_{z}$ pointing in the negative $z$ direction.
The particles have two translational degrees of freedom in the $x-$ and $z-$directions and one rotational degree of freedom about the perpendicular $y$-axes. The granular beads are described as soft disks of diameter $\sigma$,  mass $m$, and moment of inertia $I=m \sigma^2/8$, which interact via a linear contact model with viscoelastic damping between the disks and via static friction~\cite{Cundall:1979ik}. 
This model and its parameter values (See Tab.~\ref{tab}) have been chosen because they reproduce the contact properties~\cite{schaefer} of granular beads, and give results for the dynamics of an intruder in a vertically oscillated granular bed, that are in good agreement with  experiments~\cite{Sun:2006fy}.

The simulation box is driven sinusoidally, i.e., the bottom of the container is moved in time according to 
\begin{equation}
z_{b}=-\frac{1}{2} L_z + A \sin{ (\omega t)}  \ ,
\label{drive}
\end{equation}
where  $z_{b}$ is the height  coordinate of the bottom of the  container, $ A $ is the amplitude of the oscillation, $\omega$ is the frequency and $ t $ is the time. 

We traced the dynamical phase diagram  for a fixed oscillation frequency   $\omega=1.0 t_{0}^{-1}$ and  lateral wall separation $L_x/\sigma=20$. 
Reduced units are used throughout the article: the  particle mass $m$, the particle diameter  $\sigma$ and the gravitational acceleration $g$ are our fundamental units. Consequently, the derived units are the time $t_0=\sqrt{\sigma/g}$, velocity $v_0=\sqrt{g \sigma}$, force $f_0=m g$, elastic constant $k_0=m g/\sigma$ and damping coefficient $\gamma_{0}=\sqrt{g/\sigma}$.  Further details of the model are given in appendix ~\ref{aA}.

The external driving force is characterised by a dimensionless acceleration $\Gamma = A \omega^{2}/g$, corresponding to the maximum acceleration due to Eq.~(\ref{drive}) divided by the gravitational acceleration $g$.  
Alternatively, we use the dimensionless energy parameter~\cite{Pak:1993bg}
$K_{m}=\frac{A^{2} \omega^{2}}{\sigma g}$, i.e., the maximum kinetic energy per particle 'injected' in the system every period of oscillation~\footnote{This parameter is also called dimensionless shaking strength~\cite{Eshuis:980854}}.

The experiment is conducted with a monolayer of spherical  polished opaque glass beads (SiLiBeads P) with a diameter of $2\pm0.02$~mm.  
A rectangular cell made up of two glass plates $40$~mm width by $200$~mm height separated by a distance of $2.3$~mm is used to create a quasi-two-dimensional configuration. The cell is mounted on an electromagnetic shaker (Tira TV50350) with the sinusoidal frequency and amplitude controlled by a function generator (Agilent FG33220). The  acceleration is obtained by an accelerometer (Dyson 3035B2). In order to avoid the influence from electrostatic forces, the side walls of the container are made of aluminium. With a backlight illumination, the mobility of the particles are captured with a high speed camera (IDT MotionScope M3) mounted in front of the cell. The camera is externally triggered so as to take images at fixed phases of each vibration cycle. 
The snapshots captured are subjected to an image processing procedure to locate all spheres based on a Hough transformation~\cite{Kimme75}. 
Tracer particles are used to determine the  thresholds for the bouncing bed phase and for the start of convection (see appendix~\ref{aB} for details).

In order to compare the results of the simulations with the experiments, we produced results for fixed numbers of particles, namely $N=200, 400, 800, 1200, 1600$. 
The different sets are identified via the linear density $N_l=N/L_x$. This number gives an approximate value for the number of particle layers and hence, the height of the granular bed. In reality the height depends on the local structure of the granular bed, namely, the orientation of the hexagonally-ordered particles, and the type and amount of defects. 

\section{Initialisation}

In the experiment the system is initialised with a strong agitation to create a completely fluidized state, followed  by a slow ramping down of the vertical acceleration. In the simulation the initial configuration is prepared with two distinct  procedures. 
In the crystal initialisation, the particles are placed in a perfect hexagonal lattice resting at the bottom of the container. 
In the random initialisation, the particles are placed  randomly in the box. Via molecular dynamics we evolve the system until all particles have fallen under gravity and have reached a rest position.
 \begin{figure}[htdp]
\includegraphics[width=8cm]{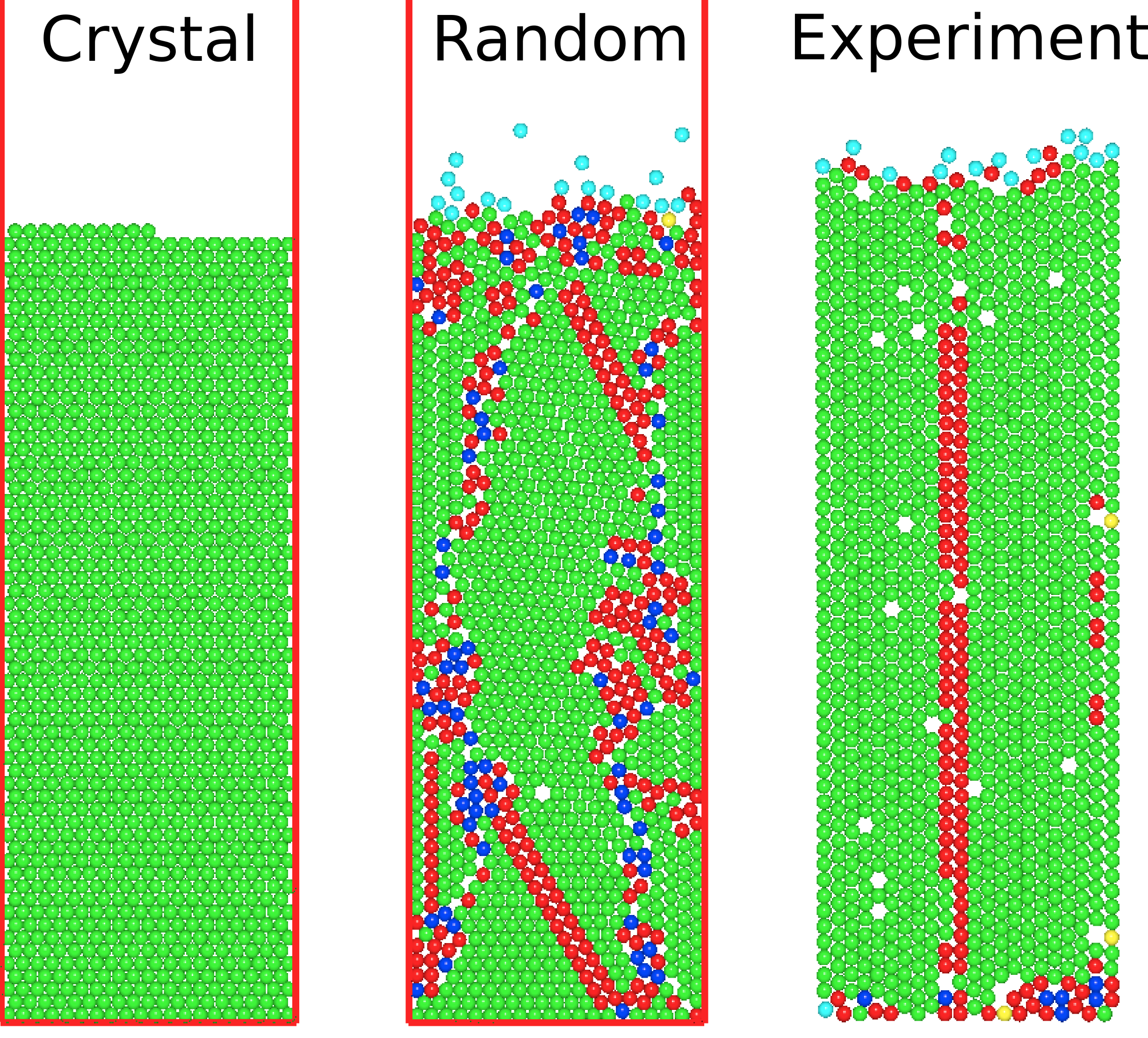}
\caption{(colour online) Simulation snapshots of the typical initial configurations for $N_{l}=60$. The colour indicate the local order of the particles according to an analysis with the $q_{6}$ order parameter~\cite{Steinhardt:1983hz}. Green (light grey) indicates hexagonal order, red (dark grey) indicates a square local order while other colours indicate a disordered configuration. From left to right: the crystalline and  random initialisations of the simulation, and  the experimental initial configuration.}
\label{snap1}
\end{figure}

\begin{figure}[htdp]
\includegraphics[width=8cm]{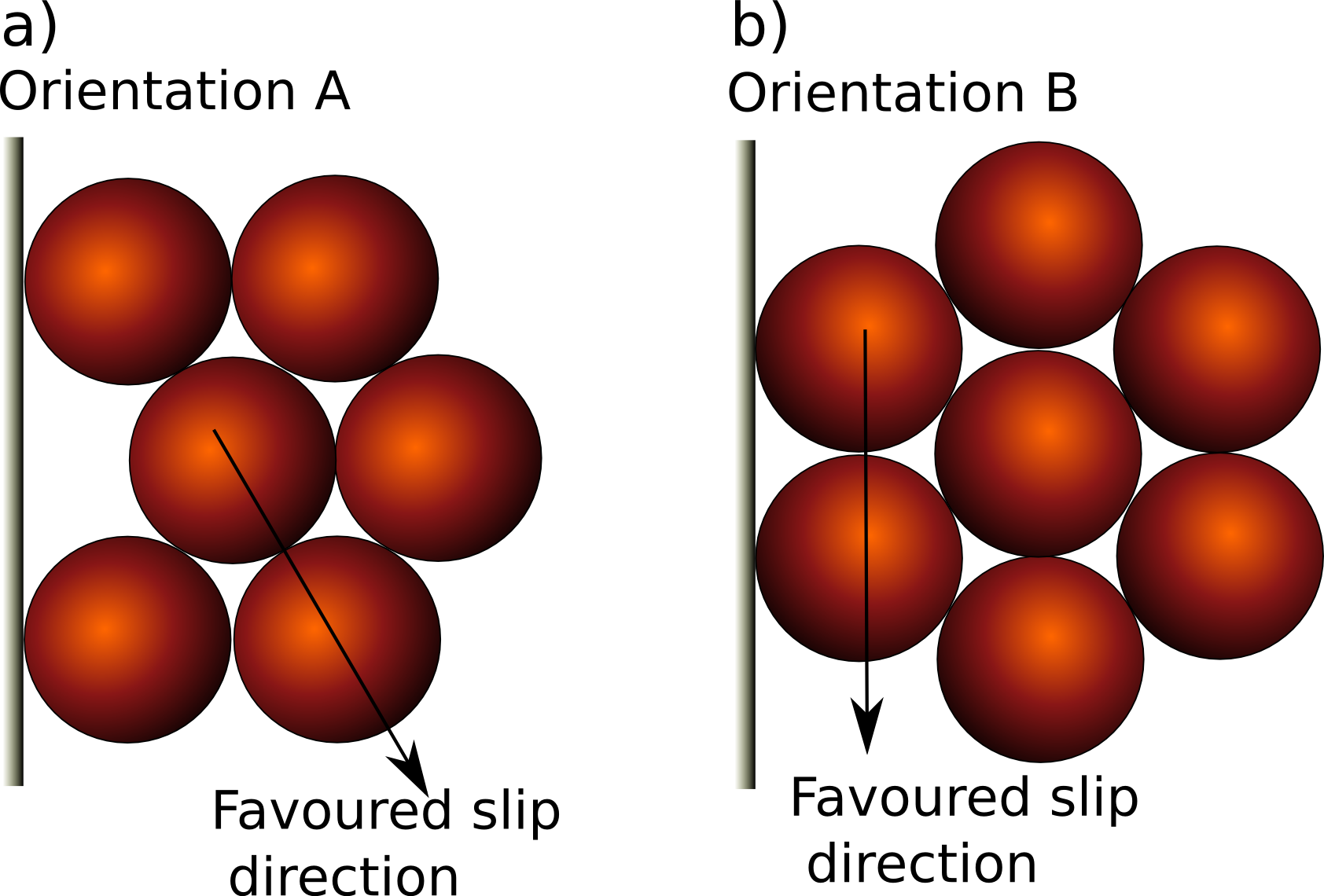}
\caption{(colour online) Sketch of the possible orientations of the hexagonal crystal in contact with the lateral wall of the container box.}
\label{hex}
\end{figure}

The resulting configurations are shown in Fig.~\ref{snap1}. We note that in both simulations initialised randomly and the experiments a large amount of particles have local hexagonal order with many topological defects, such as dislocations and grain boundaries. 
For a two dimensional system there are two favoured orientations of the hexagonal lattice, when in contact with a flat wall. In  Fig.~\ref{hex}a) the Orientation A with the [111] direction parallel to the wall is shown. In   Fig.~\ref{hex}b)  the orientation B has the [010] direction parallel to the wall. 
In the crystalline initialisation we chose to place the particles according to orientation A. After the random initialisation processes, we find  that in the experiment almost all particles have orientation B, while in simulation the orientation A seems to be favoured, but grain boundaries between the different orientations are visible. Figure~\ref{snap1} shows the configurations obtained in the three cases.

\section{Dynamical phases}

\begin{figure}[htdp]
\includegraphics[width=8cm]{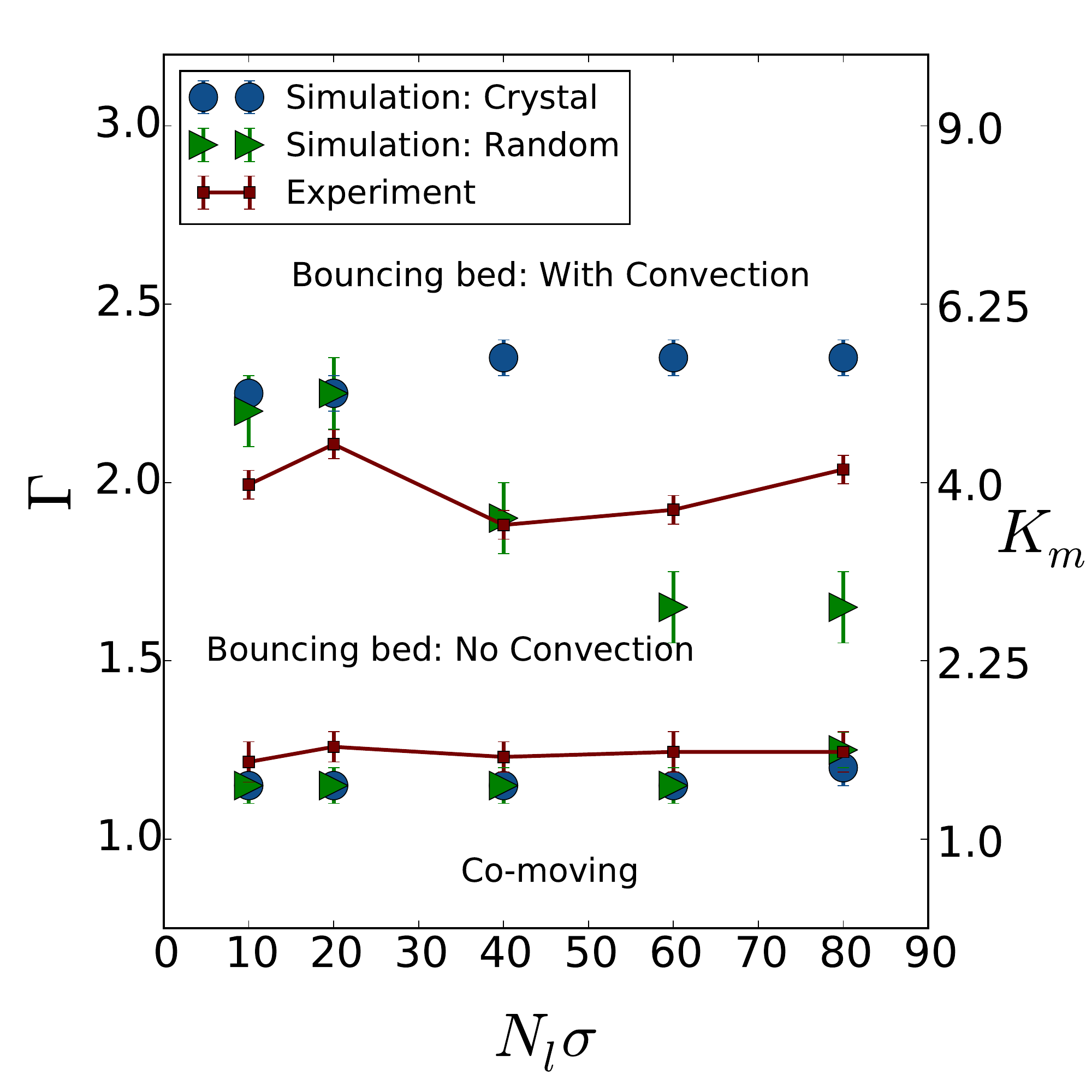}\\
\caption{(colour online) Dynamical phases at different accelerations $\Gamma$ and  linear density/number of layers $N_l \sigma$. On the right $y$-axis the energy parameter $K_{m}$ is also reported.
Symbols are for simulation results for either the system initialised randomly (triangles) or in a crystalline configuration (circles). The connected squares are the experimental results. }

\label{phd}
\end{figure}

We have systematically investigated  the onset of convection at different dimensionless accelerations/energies and heights of the granular bed. 
Figure~\ref{phd} shows the location of the dynamical phases  as a function of  the dimensionless acceleration $\Gamma$ and the energy parameter $K_{m}$ for different values of the linear density $N_l$. 
The parameter $N_l \sigma$ is also a measure of the number of layers in the granular bed, and therefore of its height.

The transition from the comoving to bouncing bed  phase occurs, as expected, at $\Gamma$ slightly larger than one, with the critical value of the acceleration increasing slightly with increasing number of layers.
We find good agreement between simulations  (green triangles/blue circles) and experiments (connected squares). In experiments, the determination of this boundary is sensitive to the accuracy of the particle position determination, which is set by the resolution of the camera. 

Inside the bouncing bed regime, we observe a transition from a bouncing bed phase without convection to a bouncing bed phase with convection. 
We find reasonable agreement between experiments and the simulations initialised randomly.
For $N_l \sigma$=60,80 the model underestimates the amount of energy necessary for the convection to start. The discrepancy is probably related to variability in the concentration of defects, as well as to the presence of the front and back wall in the experiments.
For simulations that start without topological defects we consistently detect the onset of convection (blue circles) at higher values of the acceleration with respect to the experiments and to the simulations started with defects. 
The enhancement of the convection  due to the presence of defects explains the observation of~\citet{Poschel}  that an intruder can initiate convection. The presence of an intruder  induces topological defects~\cite{deVilleneuve:2005ih} that initiate the convection and  segregation. 

\section{Onset of convection}

In crystalline materials collective particle movements occur via crystalline plane slips~\cite{Cooke:1996gg}.  The same occurs in our granular systems, but we noted some differences in slip behaviour between simulation and experiments.
In simulation the slip occur primarily along the oblique directions, while in the experiment it occurs mainly along the vertical direction. 
The difference can be explained by the different orientation of the hexagonal crystal in the two cases.  
The granular particles experience a shear stress due to the wall of the container along the vertical direction. In the experiments the  particles are oriented like in Fig.~\ref{hex}b), and slips occur preferentially in the vertical direction. On the other hand, in simulations we find more particles with an orientation like in Fig.~\ref{hex}a)  and the pyramidal slip planes will be activated first.  
As long as plane slips are concerned, the worst case scenario is represented by 
a perfect crystal with orientation A.
Since no defects are initially present, slips along the pyramidal planes can only occur if defects are nucleated first. Another difference is that in experiments we often observe single convective rolls similar to the  observation in a two-dimensional rotating cell~\cite{Rietz12_prl}. This type of convective motion is not detected in simulation and the reason for the discrepancy is likely due to a small tilt of the side walls~\cite{Knight:1993bg}.

\begin{figure}[htdp]
  \includegraphics[width=8cm]{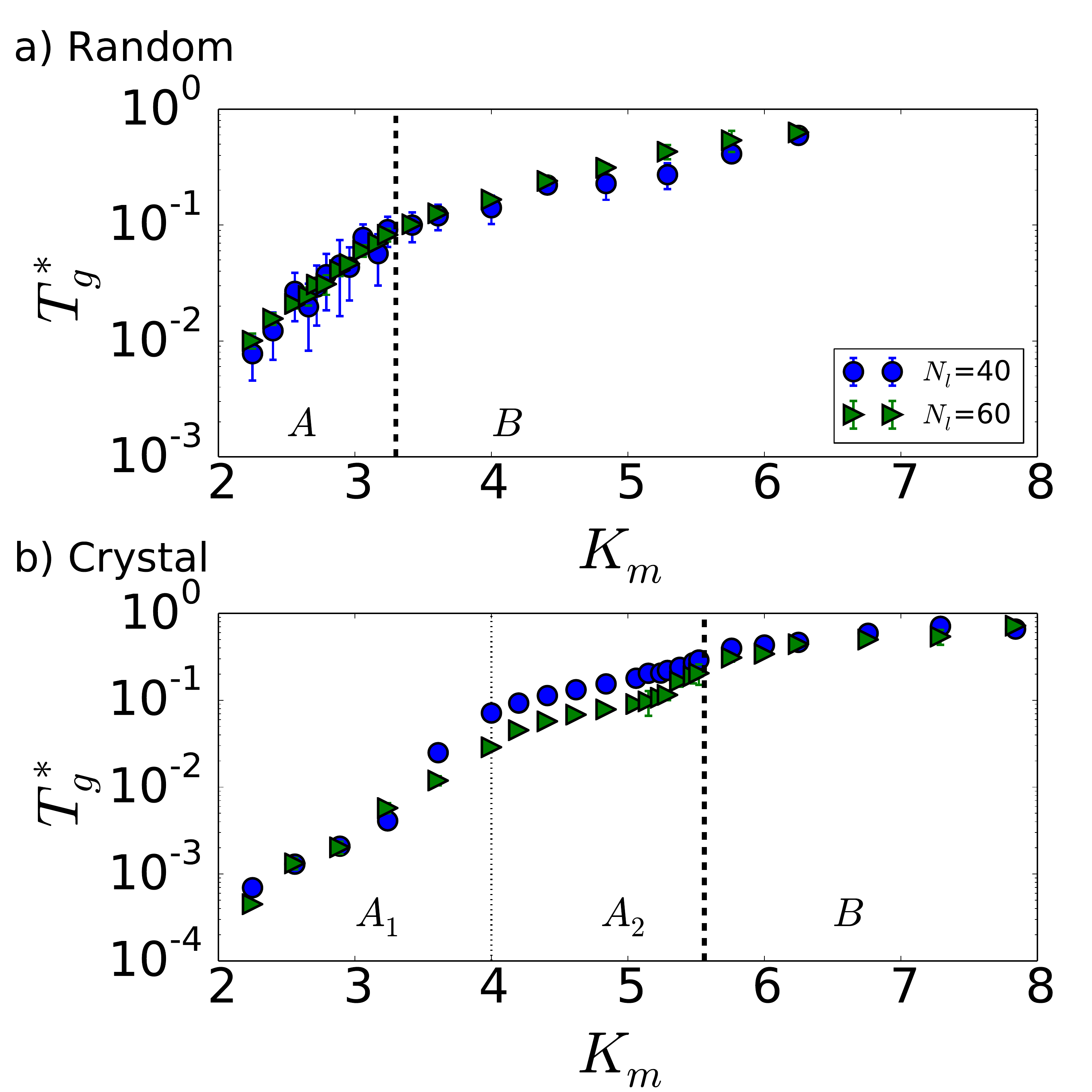}
\caption{(colour online) Granular temperature  $T_{g}$ versus dimensionless  energy parameter  $K_{m}$. The dashed lines separate regions with different slopes of the temperature curves. a) For the system initialised randomly, i.e. with defects. b) For the system initialised in a perfect hexagonal lattice, i.e. without initial defects.}
\label{temp}
\end{figure}

In order to clarify the role of defects in the onset of convection, we analyse in computer simulations the  granular temperature $T_{g}$ in relation to the average input energy $K_{m}$
The   granular temperature is defined as 
\begin{equation}
T_{g}=\frac{m}{N} \langle  \sum_{i=1}^{N} \frac{1}{2} (\vec v_{i}(t)-\vec v_{\rm cm}(t))^{2} \rangle \,
\end{equation}
where $\vec v_{\rm cm}$ is the centre of mass velocity and $\vec v_{i}(t)$ is the velocity of particle $i$ at time $t$ and $\langle \rangle$ indicates a time average.
Figures~\ref{temp}a-b show the behaviour of the reduced granular temperature $T_{g}^{*}=T_{g}/(m g \sigma)$ as a function of the dimensionless energy parameter  $K_{m}$ for  systems initialised   with and without defects, respectively.
In both cases the temperature  increases monotonically over many orders of magnitude.
The curve gives an indication of how much  input energy $K_{m}$ is converted into the kinetic energy of the granular particles. 

We divide the energy  $K_{m}$ parameter space in regions with different slopes of the   temperature curves. 
For the random initialisation (Fig.~\ref{temp}a) we find two regions: $A$ and $B$. The transition occurs  at $K_{m} \simeq 3.3$ as signalled by a change in slope of the temperature curve. Comparing the energy value for this transition with the  diagram of Fig.~\ref{phd}, we note that it  corresponds  to the energy value at the onset of convection.

On the other hand, for the case initialised without defects (Fig.~\ref{temp}b)  we can distinguish three  regions.  Following the naming convention used in Fig.~\ref{temp}a, we indicate with $B$ the region  where convection is detected. The region $A$, without convection, is now divided in two subregions $A_1$ and $A_2$. 

\begin{figure}[htdp]
    \includegraphics[width=8cm]{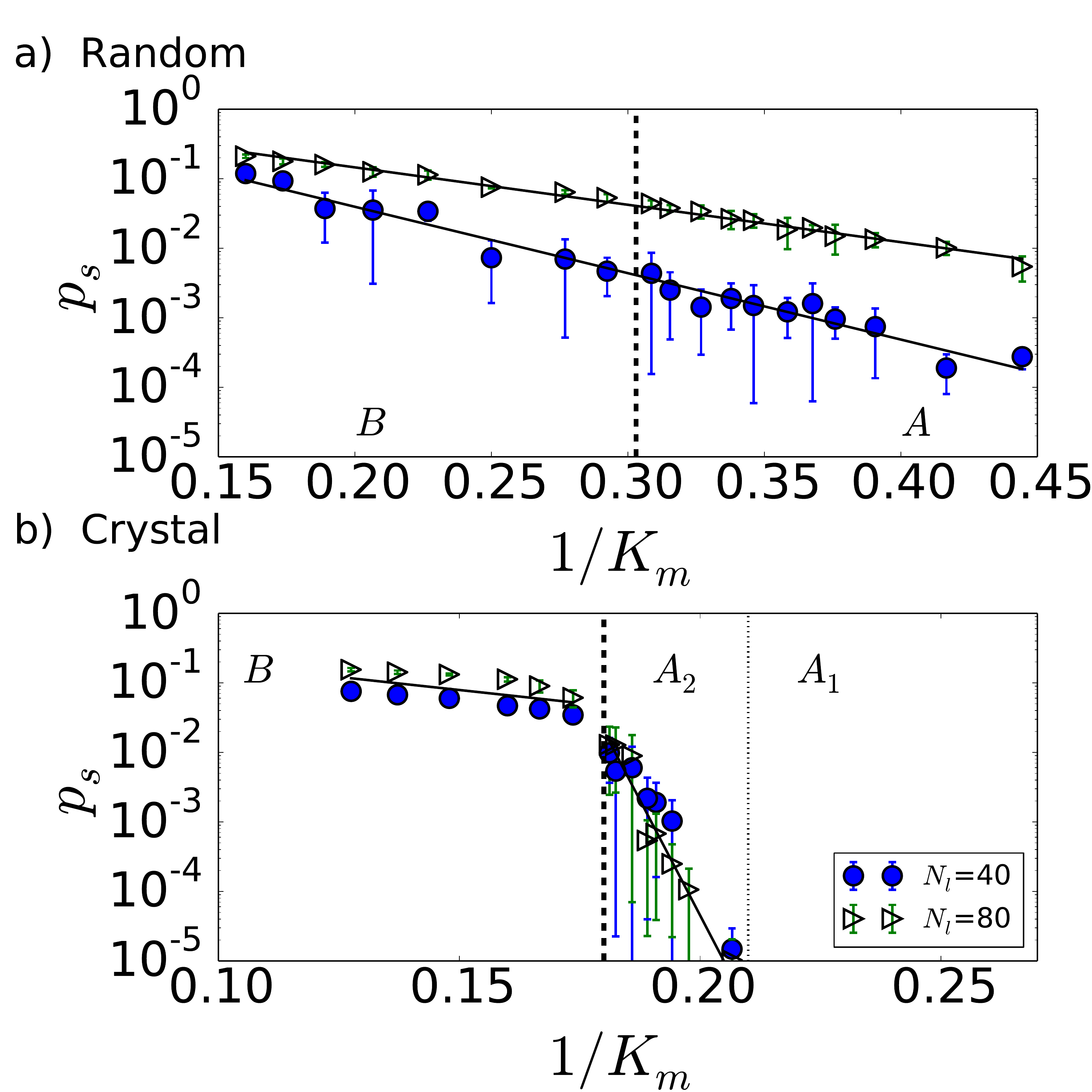}
\caption{(colour online)  Slip probability $p_{s}$ versus the inverse dimensionless energy parameter $1/K_{m}$.  The continuos lines represent exponential fit functions. The dashed lines separate the regions defined in Fig.~\ref{temp}. a) For the system initialised randomly, i.e. with defects. b) For the system initialised in a perfect hexagonal lattice, i.e. without initial defects.}
\label{Dslip}
\end{figure}

In order to clarify the origin of the  different regions of temperature behaviour, we investigate  the slip probability of a  particle in the bulk of the granular bed, i.e. away from the top free interface~\footnote{We have chosen to exclude particles closer than 10 $\sigma$ to the free interface.}.
The slip of a particle  is detected if a displacement larger than $0.2~\sigma$ is measured for at least two nearest neighbours after one oscillation. 
We define the slip probability $p_{s}$ as the fraction of particles which undergo a slip in one oscillation period. The value of $p_{s}$ is averaged over  200 periods of oscillation. 
The slip probability as a function of  $1/K_{m}$ is shown in Figs.~\ref{Dslip}a-b for the random and crystal initialisation, respectively. 
For the random case we do not observe a change in the slip behaviour between regions $A$ and $B$. On the other hand a quite dramatic change of behaviour is observed for the crystal case. 
In particular, in region $A_1$ the number of detected slip is exactly zero, while region $A_2$ is characterised by a quite steep increase of slip events, which slows down considerably upon onset of convection (region $B$).

In all cases the curves can be fitted to an Arrhenius law, $p_{s} \propto \exp(- E_{b}/K_{m})$, indicating the presence of  activating mechanisms.  
The energy parameter $E_{b}$ is a measure of the barrier height and the dimensionless energy  $K_{m}$ assumes the role of a reservoir temperature. 
For the random case just  a single curve can be fitted to the entire range of inverse energies. From the fit we find an average dimensionless barrier height $E_{b}=7\pm1$.
For the case initialised with the crystal, we find a barrier $E_{b}=130\pm 20$ in region $A_2$, and a barrier $E_{b}=7\pm 1$ in region $B$.

Since the barrier height is the same for the system with initial defects, independent of the presence of convection, we can speculate that the activation mechanism indicated by the Arrhenius equation is the activation of slip events in a crystal with topological defects.
Interestingly, the same barrier is measured for the system without initial defects in region B, suggesting the same activation mechanism. But in order for this to occur defects must first nucleate inside a perfect crystal, and we speculate that the very high barrier in region $A_2$ is due to the nucleation of defects in a perfect hexagonal crystal.

\section{Conclusions}
In conclusion, using two-dimensional computer simulations and experiments we locate the wall-induced convection in the bouncing bed region in the dynamical phase diagram of vertically vibrated  granular matter.

For the onset of convection, we find a reasonable  agreement between the experimental results and those from simulations initialised with defects.  We believe the reason behind the discrepancy is the presence of a front and a back wall in the experimental box, as well as some  variability, due to the presence of defects. 
Other phenomenological differences between simulations and experiments are observed for the favoured crystal orientation and slip events. We think that these differences are due to a small tilt of the experimental box  and small variations of the experimental box width $L_x$. 

For a system initialised in a perfect crystal, i.e., without initial defects, we consistently observe  the wall-induced convective motion to occur at higher values of the dimensionless energy parameter (shaking strength) with respect to the system initialised randomly.
From the analysis of the granular  temperature we distinguish different regions based on the slope of the temperature as a function of the dimensionless energy $K_{m}$. 
For systems initialised randomly, i.e. with many initial defects, we distinguish  two regions ($A$ and $B$) of temperature. The transition between the two regions occurs at the onset of convection, but we do not observe any change in the slip probability in the transition between regions $A$ and $B$.

On the other hand, for  systems initialised with a perfect hexagonal lattice, i.e. without initial defects, we distinguish three regions in the temperature curve.
Moreover, in this case the slip probability  has very different behaviour in the three regions. 
In particular, in region $A_1$ the number of detected slips is exactly zero, while region $A_2$ is characterised by a quite steep increase of slip events with the input energy, which slows down considerably upon onset of convection (region $B$).

The slip probability follows an  Arrhenius law, indicating the presence of  an activating mechanisms.  
The very high barrier in the region $A_2$ of the system without defects is due to  the nucleation of defects in the perfect hexagonal crystal. 
We noted that the barrier in region $B$ is the same for the system initialised with and without initial defects. In this region the behaviour of the slip probability is related to the activation of slip events.  

The results of our work provide an explanation for the onset of the convective regime in vertically oscillated granular systems and show that defects enhance onset of convection, i.e. systems with topological defects show convection at lower oscillation strengths, with respect to systems without defects. More work is needed to quantify the degree of variability for the onset of convection due to its sensitivity to the concentration and possibly types of defects.
Interestingly, since defects can diffuse out of the system when they reach the top of the granular bed,  regions of transient convective motion are possible, provided that the time scale for the defects diffusion is larger than the time scale for the nucleation of defects. 
This conjecture was not studied in this work, but represents an interesting avenue of future research.  Furthermore, we plan to study
how the size polydispersity of the granular particles changes the onset of the wall-induced convection as well as the influence of the  box size on the dynamical behaviour. Furthermore, we would like to explore the effect of shock waves~\cite{Huang:2006dw} on the dynamical behaviour of the system.
 
 \appendix
 \section{Details of the Model}
\label{aA}
\begin{figure}[htdp]
    \includegraphics[width=7cm]{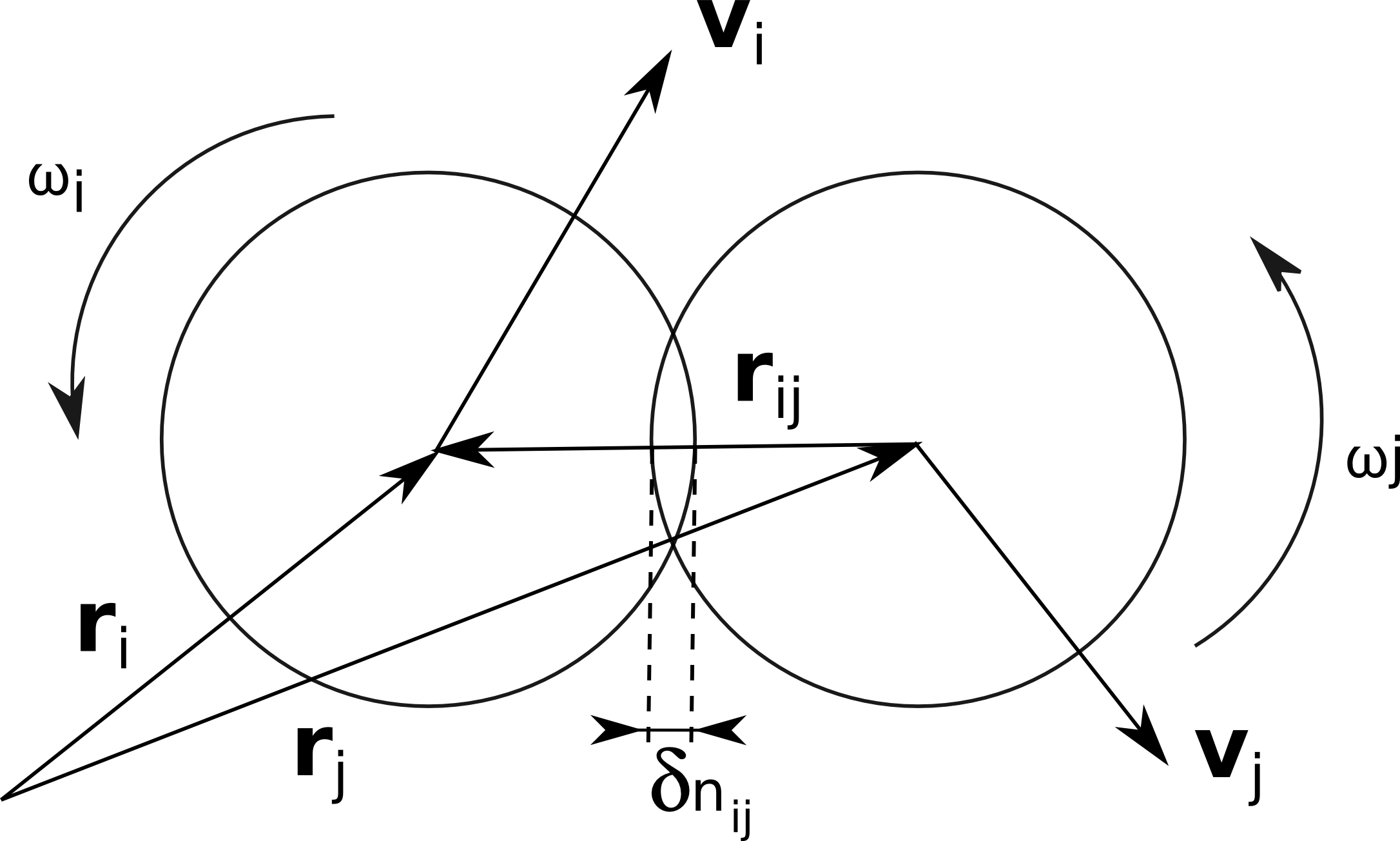}
\caption{Illustration of the contact model of two granular disks at distance $\vec r_{ij}=\vec r_i-\vec r_j$ and relative velocity $\vec v_{ij}=\vec v_i-\vec v_j$, which overlap by a distance $\delta_{n_{ij}}$. }
\label{mod}
\end{figure}

We carry out Molecular Dynamics simulations at fixed time step $dt$, for two translational degrees of freedom and one rotational degree of freedom for a system of soft disks. An illustration of the model is shown in Fig.~\ref{mod}. 
Two particles at positions $\vec  r_i$ and $\vec r_j$ with velocities $\vec v_i$ and $\vec v_j$ and angular velocities $\vec \omega_i$ and $\vec \omega_j$  define a system with an effective mass $m_{\rm eff}=m_i m_j/(m_i+m_j)$ and a normal unit vector $\vec n_{ij}=\frac{\vec  r_i-\vec  r_j}{|\vec  r_i-\vec  r_j|}$.
The tangential direction is defined as  $\vec t_{ij}=\frac{\vec v_{t_{ij}}}{|\vec v_{t_{ij}}|}$ where 
\begin{equation}
\vec v_{t_{ij}}=\vec v_{ij} - \vec v_{n_{ij}}  - \frac{1}{2} (\sigma_i \vec \omega_i+\sigma_j \vec \omega_j) \times \vec n_{ij} \ ,
\end{equation}
with $ \vec v_{n_{ij}}=(\vec v_{ij} \cdot \vec n_{ij} )\vec n_{ij}$.

We define the displacement in the two directions  $\delta_{n_{ij}}=d-|\vec  r_i-\vec  r_j|$, with $d=1/2 (\sigma_i+\sigma_j)$, and  $\delta_{t_{ij}} \vec t_{ij}= \vec v_{t_{ij}} dt$. For the disk-disk interactions we use a linear model with forces 
\begin{eqnarray}
\vec F_{n_{ij}}&=& (\kappa_n \delta_{n_{ij}} \vec n_{ij} -\gamma_n m_{eff} \vec v_{n_{ij}})\\
\vec F_{t_{ij}}&=&(-\kappa_t \delta_{t_{ij}} \vec t_{ij} -\gamma_t m_{eff} \vec v_{t_{ij}}) \ ,
\end{eqnarray}
in the normal and shear tangential  directions, respectively. The parameters $\kappa_n$ and $\kappa_t$ are the stiffness coefficients in the normal and tangential direction, respectively. The energy dissipated during the contact is regulated by the damping coefficients $\gamma_n$ and $\gamma_t$. 
In addition we model the static friction  by keeping track of the elastic shear displacement $\delta_{t_{ij}}$ over the contact lifetime and truncate it such that the Coulomb condition $| F_{t_{ij}} |< |\mu F_{n_{ij}}|$ is satisfied, where   $\mu$ is the static friction coefficient. 
The same kind of interaction is used between the particles and the container wall.
We also consider the gravitational force $\vec{g}=-  g \vec {e_z}$, where $\vec{e_z}$ is the unit vector pointing in $z$ direction.

Once the forces on all particles are known the total  force and torque $\tau_i$ on a particle $i$ is determined by
\begin{eqnarray}
\vec F_i&=&m \vec g +\sum_j (\vec F_{n_{ij}}  +\vec F_{t_{ij}} ) \\ \nonumber 
\vec \tau_i&=&-\frac{1}{2} \sum_j \sigma_j \vec n_{ij} \times \vec F_{t_{ij}} \ .
\end{eqnarray}

\begin{table}[htdp]
\begin{center}
\begin{tabular}{cc|c|c}
Coefficient &&Particle-Particle &  Wall-Particle \\
\hline
 Normal stiffness  &$k_n$ & $10^6~(k_{0})$ &  $10^6~(k_{0})$\\
  Tangential  stiffness  &$k_t$ & $10^4~(k_{0})$  & $10^4~(k_{0})$\\
 Static friction  &$\mu$ & 0.6 & 0.6\\
 Normal damping   &$\gamma_{n}$ & 100 ($\gamma_{0})$ &  100 ($\gamma_{0})$\\
 Tangential damping   &$\gamma_{t}$ & 100  ($\gamma_{0})$ & 100  ($\gamma_{0})$\\
\hline
\end{tabular}
\end{center}
\caption{Numerical values of the simulation parameters.}
\label{tab}
\end{table}%

The typical numerical values of  simulation parameters  are shown in Table~\ref{tab}. 
In this linear model it is possible to calculate the contact duration~\cite{schaefer}
\begin{equation}
t_c = \pi \left( \frac{k_n}{m_{eff}}-\frac{\gamma_n^2}{4} \right)^{-0.5} \ .
\label{t_c}
\end{equation}
In order to obtain an accurate integration of the equation of motion during contact, the time step of the simulation is chosen to be $\delta t \approx t_c / 50$~\cite{lee_forces}.

 \section{Detection of the dynamical phases}
 \label{aB}

 The detection of the convective motion, in both computer simulations and experiments, is performed by using $N_t$ tracer particles, which are initially positioned at the centre of the oscillating box in a straight horizontal line. 
In order to determine the threshold for convection, we analyse the deviation of the imaginary line connecting the tracer particles from the initial straight horizontal configuration. This is carried out by calculating, at the beginning of each cycle, the variance  the tracers height from their average height 
\begin{equation}
s^2_j=\frac{1}{N_t} \sum_{i=1}^{N_t} \left ( z^i_j - \langle z_j\rangle \right )^2 \ ,
\end{equation}
where $\langle z_j \rangle $ is the average vertical position of the tracers at frame $j$ and $z^i_j $ is the vertical position of tracer $i$ at the beginning of cycle $j$.

An average variance over all  $N_f$ frames is calculated
\begin{equation}
 s^2_{\rm c}  = \frac{1}{N_f}\sum_{j=1}^{N_f} s^2_j \ ,
\end{equation}
and the start of the convection is identified via the condition $s_{\rm c} >  \sigma$, where $\sigma$ is the diameter of the grains. 

The detection of the bouncing bed dynamical phase is performed by detecting the detachment of the granular bed from the bottom plate, which can occur at any phase of the oscillating cycle. Therefore, we calculate the average height of the tracers with respect to the oscillating plate for each phase $k$

\begin{equation}
s(k)=\frac{1}{N_f}\sum_{j=1}^{N_f}(\langle z_j(k) \rangle_i-z_{\rm b}(k)),
\end{equation}

\noindent where $\langle z_j(k) \rangle_i=\frac{1}{N_t}\sum_{i=1}^{N_t} z_j^i(k)$ is the average height of all $N_t$ tracer particles at phase $k$ and cycle $j$, $z_{\rm b}(k)$ is the height of the plate at phase $k$, and $N_f$ is the total number of frames at a certain phase.

Consequently, the variance over all $N_k$ phases is calculated

\begin{equation}
s^2_{\rm b} = \frac{1}{N_k} \sum_{k=1}^{N_k} s^2(k),
\end{equation}

and the bouncing bed phase is identified with the condition $s_{\rm b} > \sigma/30$.  The value $ \sigma/30$ corresponds to half a pixel of the experimental images. For the sake of comparison, the same value is used in the analysis of the simulation trajectories.

\begin{acknowledgments}  
The authors  thank Ingo Rehberg and Matthias Schmidt for discussions and acknowledge Philipp Ramming for the help in image processing.  Maximilian von Teuffenbach, Andreas Fischer and  Philip Krinninger are acknowledged for helping with the initial development of the granular simulation code as a part of their final bachelor projects.  K.H. is supported by the DFG through Grant No. HU1939/2-1. 
\end{acknowledgments}  

\bibliography{refs}

\end{document}